\documentclass[%
 aip,
 amsmath,amssymb,
 reprint,%
]{revtex4-1}

\usepackage{graphicx}% Include figure files
\usepackage{dcolumn}% Align table columns on decimal point
\usepackage{bm}% bold math
\usepackage{bbm}
\usepackage[utf8]{inputenc}
\usepackage[T1]{fontenc}
\usepackage{mathptmx}
\usepackage{etoolbox}
\usepackage{algorithm}
\usepackage{algpseudocode}
\usepackage{graphicx}% Include figure files
\usepackage{dcolumn}% Align table columns on decimal point
\usepackage{bm}% bold math
\usepackage{dsfont}
\usepackage[hidelinks]{hyperref}
\usepackage[capitalise]{cleveref}

\makeatletter
\def\@email#1#2{%
 \endgroup
 \patchcmd{\titleblock@produce}
  {\frontmatter@RRAPformat}
  {\frontmatter@RRAPformat{\produce@RRAP{*#1\href{mailto:#2}{#2}}}\frontmatter@RRAPformat}
  {}{}
}%
\makeatother
\begin{document}

\preprint{APS/123-QED}

\title{Stochastic Distinguishability of Markovian Trajectories}

\author{Asawari Pagare}
 \thanks{These authors contributed equally to this work and their names are ordered alphabetically.}
 \author{Zhongmin Zhang}
 \thanks{These authors contributed equally to this work and their names are ordered alphabetically.}
  \author{Jiming Zheng}
 \thanks{These authors contributed equally to this work and their names are ordered alphabetically.}
\author{Zhiyue Lu}%
 \email{zhiyuelu@unc.edu}
\affiliation{%
Department of Chemistry, University of North Carolina-Chapel Hill, NC, USA}%

\date{\today}% It is always \today, today,

\begin{abstract}
The ability to distinguish between stochastic systems based on their trajectories is crucial in thermodynamics, chemistry, and biophysics.
The Kullback-Leibler (KL) divergence, $D_{\text{KL}}^{AB}(0,\tau)$, quantifies the distinguishability between the two ensembles of length-$\tau$ trajectories from Markov processes A and B. However, evaluating $D_{\text{KL}}^{AB}(0,\tau)$ from histograms of trajectories faces sufficient sampling difficulties, and no theory explicitly reveals what dynamical features contribute to the distinguishability. This work provides a general formula that decomposes $D_{\text{KL}}^{AB}(0,\tau)$ in space and time for any Markov processes, arbitrarily far from equilibrium or steady state. It circumvents the sampling difficulty of evaluating $D_{\text{KL}}^{AB}(0,\tau)$. Furthermore, it explicitly connects trajectory KL divergence with individual transition events and their waiting time statistics. The results provide insights into understanding distinguishability between Markov processes, leading to new theoretical frameworks for designing biological sensors and optimizing signal transduction.
 
\end{abstract}

\maketitle

\section{Introduction}
Markov models are powerful in describing various phenomena in physics, chemistry, and biology \cite{gillespie1991markov, fox1978gaussian, tamir1998applications, anderson2011continuous, allen2010introduction,ge2012markov, yates2017multi, chung2013lectures, streit2006quantum, propp1985thermodynamic, mugnai2020theoretical, van2013stochastic, faggionato2009non}. These inherently probabilistic models are particularly effective in capturing the dynamic behaviors of complex systems \cite{leenders1995models, snijders2001statistical, lambiotte2014random, dixit2015inferring, cascetta1989stochastic}, and particularly non-equilibrium behaviors, including stochastic thermodynamic processes, chemical reaction kinetics \cite{anderson2011continuous, erdi2016stochastic, carrillo2016optimization, prinz2011markov, goutsias2013markovian}, and the configuration dynamics of biomolecules \cite{chodera2014markov, knoch2019non, meerbach2006multiscale, schutte2012optimal}. In biology, Markov models have provided invaluable insights into coarse-grained configuration dynamics of complex molecules, elucidating sub-cellular processes such as gene regulation networks \cite{chu2017markov, booth2005markov, zhang2008construction}, molecular motors \cite{materassi2010stochastic, faggionato2008averaging}, ion channels \cite{lampert2014markov, linaro2022markov, slowey2022sloppy}, and various cellular processes \cite{yates2017multi, bressloff2014stochastic, rengifo2002intracellular}.

The ability to distinguish between two Markov systems plays a crucial role in solving various practical problems. For example, consider identifying mutations in proteins. By examining how a protein's stochastic trajectories traversing various meta-stable states differ from that of a wild-type protein, one can tell the difference between the mutant and the wild-type. 
Another example is signal pattern recognition. Consider a sensor as a Markov system influenced by external signals. The sensor's ability to distinguish between various signal patterns is demonstrated by its ability to generate different stochastic trajectories in response to each pattern.

In this work, we adopt the master equation description of a Markov process on a finite number of states $x \in \{ 1, 2, \cdots, N \}$: 
\begin{equation}
\label{eq:master}
    \frac{{\rm d} p(x;t)}{ {\rm d} t} = \sum_{x'\neq x}R_{xx'}(t) p(x';t) + R_{xx}(t) p(x;t),
\end{equation}
where $p(x;t)$ is the probability to find the system at state $x$ at time $t$, $R_{xx'}(t)$ is the probability transition rate from state $x'$ to state $x$ at time $t$ and $R_{xx}(t)=-\sum_{x'\neq x}R_{x'x}(t)$ preserves the normalization. The master equation can also be written in a matrix form:
\begin{equation}
    \frac{\mathrm{d} \vec{p}(t)}{\mathrm{d}t} = \hat{R}\vec{p}(t),
\end{equation}
where $\vec{p}(t) = \{p(x;t)\}$ is a vector and $\hat{R} = \{ R_{x'x} \}$ is the rate matrix.

From an observational perspective, two Markov processes can be distinguished by the statistics of their states and the kinetic transitions among states (trajectories). The state transition trajectory contains more information about the Markov process than the probability distribution over the state space \cite{tang2021quantifying, PhysRevLett.131.128401, PhysRevX.13.041017,pagare2023theoretical,minas2020multiplexing}. To illustrate this, consider two Markov processes with the same stationary probability distribution, yet one system satisfies detailed balance conditions, and the other is dissipative. If both systems reach the stationary distribution, the observer cannot distinguish the two processes by merely observing the statistics of their states. However, observing the trajectories of the two processes can easily distinguish the dissipative system (with non-zero net current) from the detailed balance system (without net current). This argument has interesting indications in biological information sensing: it has been shown that the ensemble of configuration trajectories for a ligand-receptor sensor contain more information than the statistical average of its states. Moreover, by observing the trajectory of a binary-state sensor, one can simultaneously infer multiple environmental parameters (multiplexing) \cite{pagare2023theoretical, minas2020multiplexing, singh2015accurate}. 

\section{Theory}
\subsection{Distinguishability as the Kullback-Leibler Divergence }
Let us consider two ensembles of stochastic trajectories $X_{\tau}$ from two Markov processes $A$ and $B$. Each trajectory is of time length $\tau$. Here, the probability distribution of length-$\tau$ stochastic trajectories for Markov process A is denoted by 
\begin{equation}
    P^{A}[X_{\tau}] =\lim_{\delta t\rightarrow 0} p^A(x_0;0)\cdot\prod_{i=0}^{\tau/\delta t-1}p^A(x_{i+1}|x_{i};t_i),
    \label{eq:Prob_trajectory}
\end{equation}
where $p^A(x_0;0)$ is the initial state distribution of process A and $p^A(x_{i+1}|x_{i};t_i)$ is the transfer probability to state $x_{i+1}$ conditioned from state $x_i$ at time window $[t_i,t_i+\delta t)$.
Here in the limit of $\delta t\ll 1$, $X_\tau$ is a continuous-time discrete-state trajectory of time length $\tau$, and $P^{A}[X_{\tau}]$ becomes the probability density of a continuous trajectory.  
The distinguishability between the ensembles of trajectory (from time $0$ to $\tau$) observed from two Markov processes $A$ and $B$ can be characterized by the following Kullback-Leibler (KL) divergence \cite{peliti2021stochastic}:
\begin{equation}
    D_{\text{KL}}^{AB}(0,\tau) = \int P^{A}[X_{\tau}]\cdot\ln\frac{P^{A}[X_{\tau}]}{P^{B}[X_{\tau}]} ~\mathcal{D}_{X_{\tau}}.
    \label{eq:relative_entropy}
\end{equation}
Here, the value of $D_{\text{KL}}^{AB}(0,\tau)$ characterizes the distinguishability between two Markov processes from comparing ensembles of trajectories of length $\tau$. To avoid singularity of $D_{\text{KL}}^{AB}(0,\tau)$, we assume, for all $x'$ and $x$, if $R^A_{x'x}(t)\neq 0$, then $R^B_{x'x}(t)\neq 0$, and vice versa.

There are two main drawbacks in representing $D_{\text{KL}}^{AB}(0,\tau)$ merely from the above definition. Firstly, evaluation of $D_{\text{KL}}^{AB}(0,\tau)$ between two finite ensembles of trajectories suffer from insufficient sampling and slow convergence 
\cite{roldan2014estimating, martinez2019inferring}. Secondly, the definition of $D_{\text{KL}}^{AB}(0,\tau)$ in terms of the trajectory probabilities only implicitly depends on the transition kinetic rates, and thus, it misses an explicit interpretation of the distinguishability in terms of the kinetic properties of the two systems.

\subsection{Equalities for Trajectory KL Divergence}
The central result of this communication -- \cref{eq:ini_and_traj,eq:dkl0,eq:accum,eq:force-current,eq:current,eq:force} -- provide insights to understanding and evaluating $D^{AB}_{\text{KL}}(0,\tau)$, the observational distinguishability between trajectory ensembles from two Markov processes. First, $D^{AB}_{\text{KL}}(0,\tau)$ can be separated into the initial state KL divergence and the accumulated KL divergence: 
\begin{equation}
    D_{\text{KL}}^{AB}(0,\tau) = d_{\text{KL}}^{AB}(0) + D_{\text{acc}}^{AB}(0,\tau),
    \label{eq:ini_and_traj}
\end{equation}
where $d_{\text{KL}}^{AB}(0)$ denotes the state-ensemble KL divergence of the two processes' initial state distributions:
\begin{equation}
    d_{\text{KL}}^{AB}(0) = \sum_{x} p^A(x;0) \ln \frac{p^A(x;0)}{p^B(x;0)}.
    \label{eq:dkl0}
\end{equation} 
The accumulated KL divergence $D_{\text{acc}}^{AB}(0,\tau)$, is the additional distinguishability gained by observing the trajectories between $0$ and $\tau$. It follows the following decomposition in time and space (or edge):
\begin{equation}
    D_{\text{acc}}^{AB}(0,\tau) = \sum_{\text{edge } x\to x'} \int_0^ \tau    \dot D_{KL,x'x}^{AB}(t)~{\rm d} t 
    \label{eq:accum}
\end{equation}
Here, the summation $\sum_{\text{edge } x\to x'} $ includes all directed edges (connecting from state $x$ to $x'$) with non-zero rate $R_{x'x}(t)$ at time $t$.
Furthermore, we find that the distinguishability accumulation rate from each directed edge, $\dot D_{KL,x'x}^{AB}(t)$, can be written as the product of the detailed current of the edge $J^A_{x'x}$ and a divergence force $F^{AB}_{x'x}(t)$ (see SI.I):  
\begin{equation}
    \dot D_{KL,x'x}^{AB}(t) = J^A_{x'x}(t) \cdot F^{AB}_{x'x}(t) \geq 0,
    \label{eq:force-current}
\end{equation}
The detailed probability current from $x$ to $x'$ at time $t$ is 
\begin{equation}
    J^A_{x'x}(t)=R^{A}_{x'x}(t) \cdot p^A(x;t)  \geq 0.
    \label{eq:current}
\end{equation}
The divergence force itself is a KL divergence between two transition dwell time distributions: 
\begin{align}
    F^{AB}_{x'x}(t) &= D_{\text{KL}}[p_w^{A,x'x,t}(s)||p_w^{B,x'x,t}(s)]  \nonumber \\ 
    &= \ln {\frac{R^{A}_{x'x}(t)}{R^{B}_{x'x}(t)} + \frac{R^{B}_{x'x}(t)}{R^{A}_{x'x}(t)} -1} \geq 0,
    \label{eq:force}
\end{align}
where $p_w^{A,x'x,t}(s) = R^{A}_{x'x}(t) \exp\left (-R^{A}_{x'x}(t) \cdot s\right)$ is the exponential distribution of dwell time $s$ in state $x$ before a transition to state $x'$ occurs, characterized by the transition probability rate $R^{A}_{x'x}(t)$ at time $t$.

%The decomposition of $D^{AB}_{\text{KL}}(0,\tau)$ reveals several critical properties of the distinguishability between Markov systems. Moreover, the theory sheds light on efficient sampling methods to evaluate $D^{AB}_{\text{KL}}(0,\tau)$ from limited samples of trajectories:

\subsection{Key Properties of Trajectory KL Divergence}

The results reveal several critical properties of the trajectory KL divergence, $D^{AB}_{\text{KL}}(0,\tau)$. It provides the three perspectives to understand the distinguishability between two Markov systems.

\subsubsection{The initial and the accumulated divergence} When the observed trajectory length reduces to $\tau \to 0$, $D_{\text{KL}}^{AB}(0,\tau) = d^{AB}_{\text{KL}}(0)$. In this limit, the trajectory KL divergence of the two Markov processes reduces to the KL divergence between the two initial state distributions. That is, by observing infinitely short trajectories, the distinguishability of two Markov processes equals the distinguishability between the systems' initial state distributions.
For any finite-time observation $\tau>0$, \cref{eq:force-current} implies that as trajectory length $\tau$ increases, the distinguishability always accumulates with a non-negative rate: $\dot D^{AB}_{KL,x'x}(t) \geq 0,~\forall \{(t,x,x') | x' \neq x\}$. In other words, the longer the observed trajectory, the greater the revealed distinguishability between the two Markov processes. It is noteworthy that the initial state distribution of process B does not impact $D_{\text{acc}}^{AB}(0,\tau)$, since its effect is fully captured by the initial $d^{AB}_{\text{KL}}(0)$.

\subsubsection{Temporal decomposition of $D_{\text{KL}}^{AB}(0,\tau)$} Here, we define the temporal additivity of the trajectory KL divergence by decomposing one time period into two consecutive segments:
\begin{equation}
    D_{\text{KL}}^{AB}(0,\tau) = D_{\text{KL}}^{AB}(0, t^*) +D_{\text{KL}}^{AB}(t^*, \tau) - d_{\text{KL}}^{AB}(t^*),
    \label{eq:addative}
\end{equation}
where $D_{\text{KL}}^{AB}(t_a, t_b)$ denotes the trajectory KL divergence for time starting at $t_a$ and ending at $t_b$, and the lowercase $d_{\text{KL}}^{AB}(t^*)$ denotes the state-ensemble KL divergence between two systems evaluated at time $t^*$. This additivity formula indicates that one can obtain the KL divergence of trajectories starting at $0$ and ending at $\tau$ as the summation of that for two (or more) consecutive trajectories $(0,t^*)$ and $(t^*,\tau)$, after the removal of the overlapping term(s) $d_{\text{KL}}^{AB}(t^*)$. 
This decomposition can also be represented in a cleaner form in terms of accumulated KL divergence:
\begin{equation}
    D_{\text{acc}}^{AB}(0,\tau) = D_{\text{acc}}^{AB}(0, t^*) +D_{\text{acc}}^{AB}(t^*, \tau).
    \label{eq:addative_acc}
\end{equation}

\subsubsection{Spatial (edge-wise) decomposition of $D_{\text{KL}}^{AB}(0,\tau)$} \cref{eq:accum,eq:force-current} reveal the non-negative spatial (edge-wise) additivity of the trajectory KL divergence, where accumulated KL divergence is the summation of non-negative contributions from each directed transition via connected edges. Furthermore, \cref{eq:force-current} shows that the contribution from each directed edge equals the product of the detailed probability flow rate \cref{eq:current} and a force-like weighting factor \cref{eq:force}. The force-like factor characterizes the distinction between the two transition rates of the same edge from two Markov processes. This additivity of non-negative contribution from each edge implies that coarse-graining of states or ignoring any edge of the graph may cause underestimation of the trajectory KL divergence, which agrees with intuition. Importantly, this spatial decomposition in state transitions leads to a practical formula described at the end of this paper, allowing for the development of enhanced transition event sampling methods towards the efficient evaluation of $D_{\text{KL}}^{AB}(0,\tau)$. 

\subsection{Connection to Thermodynamic Entropy}
Entropy plays a central role in nonequilibrium thermodynamics and is closely related to the Jarzynski equality \cite{jarzynski1997nonequilibrium}, Crooks fluctuation theorem \cite{crooks1999entropy}, and other forms of fluctuation theorems \cite{seifert2008stochastic,martinez2019inferring,maes2020response}. Apparently, the temporal and spatial decomposition of $D^{AB}_{\text{KL}}(0,\tau)$ resembles the decomposition of the entropy production in stochastic thermodynamics \cite{seifert2008stochastic}.
However, the connection between $D^{AB}_{\text{KL}}(0,\tau)$ and the entropy production is non-trivially dependent on the construction of Markov process $B$.

In stochastic thermodynamics, the entropy production of a Markov process can be obtained from the logarithm ratio of the probability measure of a trajectory and the probability measure of the time reversal trajectory, which is known as the Radon-Nikodym derivative (RND)\cite{frater1989rare}.
In contrast, the $D^{AB}_{\text{KL}}(0,\tau)$ defined in this work involves the RND of one trajectory's probability measures from two Markov processes $A$ and $B$. 
The RND and generalized fluctuation theorems under various choices of conjugate processes can be found in \cite{ge2021martingale,dechant2020fluctuation,horowitz2010nonequilibrium,ge2012markov, chetrite2011two,chetrite2008fluctuation,martinez2019inferring,maes2020response,gallavotti1995dynamical,baiesi2015inflow,chetrite2009fluctuation}. 
Notably, the trajectory KL divergence under arbitrary conjugate process was discussed for diffusive and Fokker–Planck (FP) dynamics in \cite{ge2021martingale,dechant2020fluctuation}. It is worthy of further study to determine if the results of this paper, for discrete-state Markov processes, can be directly extended to FP dynamics.

One may naturally speculate that $D^{AB}_{\text{KL}}(0,\tau)$ is related to thermodynamic entropy production, if the process $B$ is chosen as the ``reversal'' of the process $A$. However, we find that the process $B$ should be defined without inverting the direction of time or control protocol. Rather, $D^{AB}_{\text{KL}}(0,\tau)$ becomes thermodynamic entropy when process $B$ is chosen to be the pseudo-transpose of $A$, denoted by $A^*$:
\begin{equation}
    \begin{cases}
        R^{A^*}_{xx'}(t) = R^A_{x'x}(t), &\forall (x, x'), ~ x\neq x', \vspace{1em}\\
        R^{A^*}_{xx}(t) = -\displaystyle\sum_{x'\neq x} R^{A^*}_{x'x}(t), &\forall x, 
    \end{cases}
\end{equation}
where for $A^*$, the transition rate from state $x$ to $x'$ at time $t$ is chosen to be identical to the rate from $x'$ to $x$ of the the process $A$ at time $t$. The normalization is preserved for the process $A^*$ due to the choice of its diagonal elements.
Then, by assuming that the Markov process A is immersed in a heat reservoir, the time-accumulated KL divergence between processes $A$ and $A^*$ equals the entropy change of the reservoir caused by the process $A$: (see SI.II)
\begin{equation}
    D^{AA^*}_{\text{acc}}(0,\tau) = \frac{1}{k_\text{B}} \langle \Delta s^{\text{res}}[X_\tau] \rangle^A,
\end{equation}
where $k_\text{B}$ is the Boltzmann's constant, $s^{\text{res} }[X_\tau]$ is the stochastic entropy production in the reservoir due to trajectory $X_\tau$, and the angular bracket $\langle \cdot \rangle^A$ is the trajectory ensemble average for process $A$ \cite{seifert2008stochastic}.

To summarize, for a Markov thermodynamic process A and its ``conjugate'' dynamics $A^*$, when the two systems are prepared at the same initial distribution ($d_{\text{KL}}^{AA^*}(0)=0$), their trajectory KL divergence equal the reservoir's entropy change due to process A.

\section{Discussions and Applications}

In the remaining, we discuss the practical implications of the theory. We first examine how the initial state distribution affects the observational distinguishability between two Markov systems (see \cref{discuss_1,discuss_2}). Then we discuss how $D_{\text{KL}}^{AB}(0,\tau)$ responds to control protocols (input signal), and illustrate its implications for biological or artificial stochastic sensors. Specifically, the theory inspires the design of optimal sensor that can discern input signals (see \cref{discuss_3}), and reveals the optimal control protocol to discern two apparently similar systems (see \cref{discuss_4}). In the end (see \cref{discuss_5,discuss_6}), we show that the theory provides multiple new approaches to efficiently evaluate the trajectory KL divergence. These approaches get around the sufficient sampling difficulty of evaluating  $D_{\text{KL}}^{AB}(0,\tau)$ from a finite number of trajectories. 

\subsection{Distinguishability of Initial State Distributions}
\label{discuss_1}
Consider two stochastic systems that evolve according to identical kinetic rates, $\hat R^A(t)=\hat R^B(t)~\forall t$, and they differ only by their initial-state distributions, $\vec p^A(0) \neq \vec p^B(0)$. Does measuring their trajectories provide more information than the initial states regarding distinguishing the two systems? No, we show that the difference in their initial state distributions fully captures the distinguishability of the two Markov processes: From \cref{eq:force}, the divergence force is zero, $F_{x'x}^{AB}(t)=0$, for all times and edges due to the identical kinetic rates $\hat R^A(t)=\hat R^B(t)$. Then, according to \cref{eq:accum} and \cref{eq:force-current}, the accumulated KL divergence by observing longer trajectory always equals 0. Thus, \cref{eq:ini_and_traj} reduces to $ D_{\text{KL}}^{AB}(0,\tau) = d_{\text{KL}}^{AB}(0) $. In other words, additional observations of the trajectories beyond the initial state cannot improve the observational distinguishability between Markov systems with fully identical kinetic rates. An alternative proof by using the data processing inequality \cite{cover1999elements} is shown in the SI.III.

\subsection{Optimal Initial Distribution to Maximize Trajectory Distinguishability}
\label{discuss_2}
\begin{figure}[htbp]
    \centering
    \includegraphics[scale=0.58]{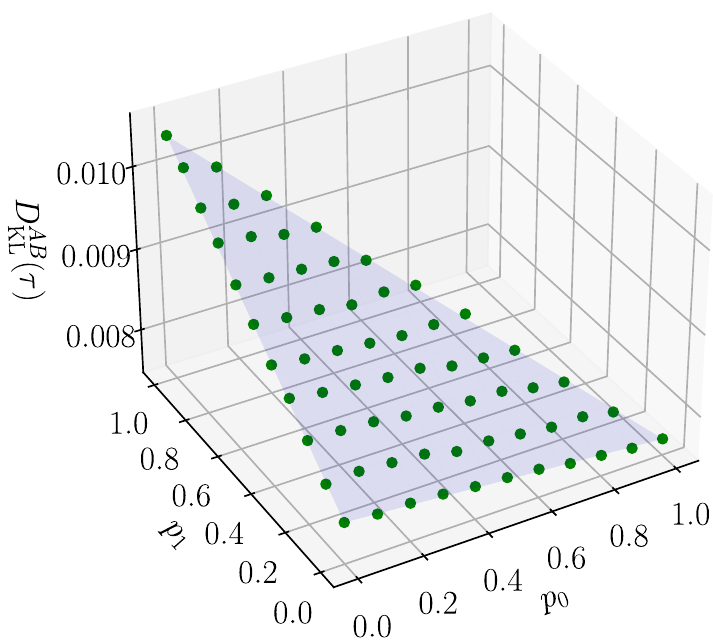}
    \caption{Initial distribution dependence of the trajectory KL divergence. Trajectories are obtained from two 3-state Markov processes with common initial distribution $(p_0,p_1,p_2)$, where $p_2=1-p_0-p_1$. Each data point represents a distinct initial distribution, and points lives on a flat plane in accordance with \cref{eq:acc_decomp}. The maximum divergence is achieved at a delta-function initial distribution $(p_0,p_1,p_2)=(0,1,0)$. See SI.IVB for details.}
    \label{fig:mpemba}
\end{figure}

Consider two Markov systems prepared at identical initial state distributions $p^A(x;0)=p^B(x;0)=p(x;0)$, yet they evolve with different kinetic rates which could be time-dependent or time-independent. How to prepare the initial distribution $p(x;0)$ of the two systems such that their trajectories most prominently reveal their distinguishability? We find that the optimal initial distribution to maximize the distinguishability is typically a delta-function distribution. In other words, there exist one (or a few degenerate) optimal state(s), from which the observational distinguishability of the two systems is maximized. To show this, recognize that the accumulated KL divergence follows an initial-state decomposition relation: 
\begin{equation}
    D_{\text{acc}}^{AB}(0,\tau) = \sum_{x} p^A(x;0) \cdot \left. D_{\text{acc}}^{AB}(0,\tau)\right|_{x_{\text{ini}}=x}
    \label{eq:acc_decomp}
\end{equation}
where $\left. D_{\text{acc}}^{AB}(0,\tau)\right|_{x_{\text{ini}}=x}$ is the accumulated trajectory KL divergence if both systems are initially prepared perfectly on a same state $x$. This result indicates that the trajectory KL divergence is linearly dependent on the initial probability distribution, as illustrated by a tilted flat plane in \cref{fig:mpemba}. See proof of the decomposition relation in SI.IVA.
As a result, the highest accumulated KL divergence must be achieved when the initial distribution is a delta function at the optimal state $x^*$, where $\left. D_{\text{acc}}^{AB}(0,\tau)\right |_{x_{\text{ini}}=x^*} \geq \left. D_{\text{acc}}^{AB}(0,\tau)\right|_{x_{\text{ini}}=x}$ for any state $x$. Very rarely, when there are degenerate optimal states $\{x^*\}$, any initial distribution purely comprising these degenerate states also gives the optimal distinguishability. 
Intuitively, from \cref{eq:accum,eq:force-current}, this optimal initial state to distinguish two time-independent Markov processes is the one that evolves in time to exhibit the largest accumulated weighted flow of transitions, with the weight defined by \cref{eq:force}. Notice that the specific choice of the optimal state depends on both the systems' kinetic property and the length of the observational time $\tau$. For illustrative purpose, consider two time-independent Markov processes with anomalous relaxations \cite{lu2017nonequilibrium,klich2019mpemba} where one finds a large separation of the speed of relaxation (big gap in the eigenvalues of $\hat R$), we find that the optimal initial state $x^*$ can be determined by observing the slowest relaxation eigenmode (see this example in Fig.~\cref{fig:mpemba} and SI.IVB).

Now, we conclude the discussion of the initial state distribution and shift the focus to temporal control protocol(s) governing/applied to two Markov systems. Here, we describe a sensor under influence of a signal $\lambda(t)$ by a Markov system $\hat R(\lambda(t))$.
Under this construction, the results of this work can be used to describe temporal pattern recognition by a sensor (see \cref{discuss_3}). It can also help the search for the optimal external control protocol to maximally reveal/enhance the distinction between two apparently similar Markov systems with slight differences in their kinetic properties (see \cref{discuss_4}).

\subsection{Design Principle: Sensor for Discerning Temporal Patterns}
\label{discuss_3}
Various biological sensors can improve their sensing capability by utilizing non-trivial stochastic trajectories of their internal states evolving under complex external input signals \cite{hopfield1974kinetic, qian2006reducing,
minas2020multiplexing, tang2021quantifying, PhysRevLett.131.128401, PhysRevX.13.041017, pagare2023theoretical}.
Consider a stochastic sensor as a Markov system hopping between different internal states. Its time-dependent rate matrix is $\hat R(\lambda)$, and the protocol $\lambda (t)$ represents the external signal. The downstream sensory network, which interacts with the sensor but does not directly interact with the input signal, can discern two temporal protocols ($\lambda^A(t)$ and $\lambda^B(t)$) only by ``observing'' the sensor's state paths (trajectories). In this case, the maximum information the downstream sensory network can acquire is limited by the information contained within the observed ensemble of sensor's state trajectories. Thus, the trajectory KL divergence defined in this work captures the upper bound of the distinguishability of the two signals transmitted via the sensor, and an optimal sensor is one with the highest upper bound given the design restrictions.  
As a result, the theory in this paper can be used in various practical applications for improving the sensor's sensitivity or optimizing the ability of pattern recognition. This paper does NOT aim to provide a universal analytical solution for the optimal sensor's design, because the solution depends on various specific details of the practical problem. E.g., the types of protocols that need to be discerned, the physical coupling between the sensor and signal, as well as other limitations that define the available form of the sensor's rate matrix $\hat R(\lambda)$. 
Although not discussed in this paper, we have recently applied this theory to designing optimal concentration sensors and sensors capable of recognizing patterns of time-varying signal-molecule concentrations.

\subsection{Optimal Control Protocol to Reveal Differences between Markov Systems}
\label{discuss_4}
A conjugate problem of \cref{discuss_3} is to design the optimal control protocol that maximally reveals the difference between two Markov systems. To illustrate the problem with a practical example, consider two proteins with subtle differences: a wild type and a mutant. The two molecules share the same number of metastable configurations (states); however, the transition rates between states are different for the two, and their rates are influenced by an external control parameter $\lambda$ (e.g., electric field, temperature, or pH). If the two molecules are subtly different in their kinetic rates and/or their rates' dependence on the external control, merely observing the steady-state distribution of the molecules in a stationary environment may reveal, to some extent, the differences between the two molecules. However, our theory indicates that one can enhance the ability to distinguish the two systems when we drive them out of the steady state by the same control protocol $\lambda(t)$ and observe the difference in their stochastic trajectory ensembles (see \cref{eq:relative_entropy}).
Moreover, the theory allows for the search of the optimal control protocol $\lambda^*(t)$ to maximally reveal the difference between the two systems. Given $\hat R^A(\lambda)$ and $\hat R^B(\lambda)$, the specific forms of the two molecule's kinetic rates and their dependence on the external control parameter $\lambda$, the search for the optimal protocol $\lambda^*(t)$ becomes a variational optimization problem of $D_{\text{acc}}^{AB}(0,\tau)$. Intuitively, the decomposition relations imply that the optimal driving protocol maximizes the accumulated number of transitions over the edges with accumulation weight factor $F_{xx'}^{AB}(t)$.

In the following, we discuss how the decomposition relation can circumvent the sampling convergence difficulties in evaluating the trajectory KL divergence. 

\subsection{Slicing Trajectory Approach for Evaluating $D_{\text{KL}}^{AB}(0,\tau)$}
\label{discuss_5}
\begin{figure}
    \centering
    \includegraphics{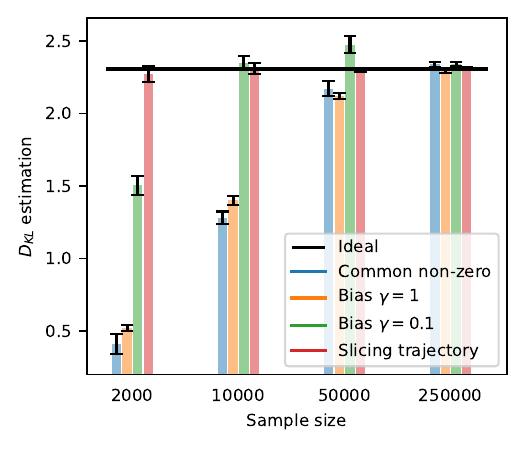}
    \caption{Trajectory KL divergence between two binary-state systems with rates $R^A_{21}=1,~R^A_{12}=2$ and $ R^B_{21}=2,~R^B_{12} = 1$. Each trajectory is 10-step long with time step equals unity. The KL divergence is evaluated from trajectory ensembles of sizes 2000, 10000, 50000, and 250000 by different methods. Each result is averaged over 10 repetition of calculations for randomly generated trajectories, with the error bar showing $\pm 1$ standard deviation from the 10 repetitions. The trajectory slicing method slices trajectory into consecutive segments of length 2. See SI.V for details.}
    \label{fig:enhanced}
\end{figure}

When directly evaluated from the definition, the convergence of $D^{AB}_{\text{KL}}(0,\tau)$ can face significant sampling difficulty. It requires that each trajectory observed in process $A$ must also be observed sufficiently many times in process $B$. This convergence difficulty grows rapidly with the trajectory length $\tau$, making it almost impossible to directly calculate the $D^{AB}_{\text{KL}}(0,\tau)$ for long trajectories. However, the temporal decomposition shown in \cref{eq:addative} makes it possible to get around the difficulty by the following:
By dividing the trajectory into multiple consecutive segments, it is easier to evaluate the segment trajectory KL divergences and reconstruct the whole-trajectory KL divergence. The desired whole-trajectory KL divergence can be obtained by adding the segment KL divergences together minus the state distribution KL divergences $d^{AB}_{KL}(t^*)$'s evaluated at the intersecting time points. This approach significantly reduces the sufficient number of trajectories needed to evaluate $D_{\text{KL}}^{AB}(0,\tau)$ without singularity. In comparison to traditional approaches, this new trajectory-slicing approach significantly reduces the number of trajectories required for an converging and accurate estimation of the trajectory KL divergence. (See Fig.~\ref{fig:enhanced} and SI.V for details.)

\subsection{Event-counting Estimation of $D_{\text{KL}}^{AB}(0,\tau)$}
\label{discuss_6}
The method proposed above still relies on the definition of KL divergence, \cref{eq:relative_entropy}, and thus, it explicitly depends on the probability density functions (histograms) of segmented trajectories. Below, we show that with merely the sets of transition events, one can evaluate the trajectory KL divergence. Specifically, the temporal and spatial decomposition relations, \cref{eq:force-current,eq:current,eq:force} allow us to express the trajectory KL divergence as a weighted statistical average of transition events:
\begin{equation}
     D_{\text{KL}}^{AB}(0,\tau) = d_{\text{KL}}^{AB}(0) + \left\langle \sum_k F_{x'_k x_k}^{AB}(t_k) \right\rangle^A ~,
     \label{eq:simu}
\end{equation}
where $k$ is the index of the transitions (from state $x_k$ to $x'_k$) within a trajectory $X_{\tau}$, and the angular bracket is an average among all trajectories obtained from process A. Practical applications of the above formula may fall into one of the three scenarios below. 

\begin{figure}
    \centering
    \includegraphics{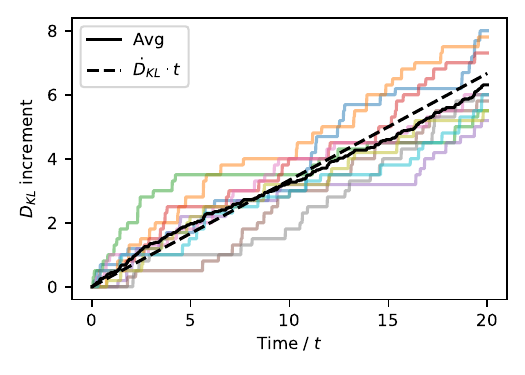}
    \caption{Evaluation of $\dot D_{KL}^{AB} \cdot t$ based on event-counting method. Trajectories are generated from two binary-state Markov systems with rates $R^A_{21}, R^A_{12}, R^B_{21}, R^B_{12} = 1, 2, 2, 1$. Color curve is accumulated $\dot D_{KL}^{AB} \cdot t$ evaluated from a single trajectory. Black solid curve is the average over the ten color trajectories. Black dashed line is the analytical result for $\dot D_{KL}^{AB} \cdot t$.}
    \label{fig:event}
\end{figure}

{\it Scenario 1:} If the transition rate matrices $\hat R^A(t)$ and $\hat R^B(t)$ are known, then one can directly evaluate the trajectory KL divergence with the above equation, by plugging in the values of $F_{x'x}^{AB}(t)$ (see \cref{fig:event}).

{\it Scenario 2:} If the transition rate matrices are time-homogeneous yet unknown, one can utilize existing tools \cite{kenney2023thermodynamically,dixit2014inferring,gotz2022blind,barendregt2023adaptive} to evaluate the transition rates for processes $A$ and $B$ and then obtain $F_{x'x}^{AB}$ to perform the calculation described in {\it Scenario 1}.

{\it Scenario 3:} If the transition rate matrices are unknown and are time-dependent, one may need to utilize the waiting time statistics of transition between state $x$ and $x'$ at each time $t$, to first evaluate the time-dependent rate matrices, and then obtain $F_{x'x}^{AB}(t)$ for calculating the trajectory KL divergence. In this case, utilizing the method proposed in \cref{discuss_5} may be more efficient for evaluating the trajectory KL divergence. 

\section{Conclusion}
In conclusion, this study rigorously delineates the stochastic distinguishability between two arbitrary Markov processes, quantified through the trajectory KL divergence, $D_{\text{KL}}^{AB}(0,\tau)$. Our development of a general formula to decompose $D_{\text{KL}}^{AB}(0,\tau)$ in both spatial and temporal dimensions not only unveils the dynamical traits contributing to the distinguishability of stochastic systems but also allows for the efficient evaluation of this divergence.

Given the flexibility in selecting Markov processes A and B, our theory is poised for broad application across various fields. It promises to extend the scope of fluctuation theorems from stochastic thermodynamics to encompass a wide range of non-thermal processes. Moreover, by analyzing trajectories timed by different mechanisms, our theory applies to clock calibration and offers insights into how a deterministic conception of time may arise from purely stochastic dynamics. Lastly, by leveraging this theoretical framework to quantify information within stochastic trajectories, we lay the groundwork for innovative design principles for biological and artificial sensors to recognize and interpret temporal patterns in input signals.

\section*{Supplementary Information}
Technical derivations that are too lengthy to be included in the manuscript and numerical simulation codes can be found in the Supplementary Materials.

\section*{Data Availability Statement}
Data sharing is not applicable to this article as no new data were created or analyzed in this study.

\begin{acknowledgments}
National Science Foundation Grant DMR-2145256 financially supports this work. ZL is grateful for constructive comments on this work from Prof. Hong Qian at the University of Washington Seattle, Prof. Ying Tang at Beijing Normal University, and Mr. Ruicheng Bao at University of Science and Technology of China.
\end{acknowledgments}

\bibliography{aipsamp}% Produces the bibliography via BibTeX.

\end{document}